\documentclass[%
 reprint,
 amsmath,amssymb,
 aps,
]{revtex4-2}

\usepackage{hyperref}
\usepackage{placeins}
\usepackage{graphicx}
\usepackage{dcolumn}
\usepackage{bm}

\begin{document}

\preprint{APS/123-QED}

\title{Statistical properties of non-flow correlations in  pp and heavy-ion collisions at RHIC energies\\}

\author{Satya Ranjan Nayak$^{1}$}
 \email{satyanayak@bhu.ac.in}

 \author{Akash Das$^{2}$}
 \email{24pnpo01@iiitdmj.ac.in}
\author{B. K. Singh$^{1,2}$}%
 \email{bksingh@bhu.ac.in}
 \email{director@iiitdmj.ac.in}
\affiliation{$^{1}$Department of Physics, Institute of Science,\\ Banaras Hindu University (BHU), Varanasi, 221005, INDIA. \\
$^{2}$Discipline of Natural Sciences, PDPM Indian Institute of Information Technology Design \& Manufacturing, Jabalpur-482005, India\\
}
\date{\today}

\date{\today}
\begin{abstract}
In this work, we have studied the two-particle cumulant in pp, d-Au, and Au-Au collisions. The two-particle cumulant was treated as an event-by-event distribution, and its skewness and kurtosis were analyzed. The non-flow correlations, like jets and decays, constantly produced a skewed distribution, regardless of the model used. On the contrary, HYDJET++ produced a smooth Gaussian distribution at higher $\eta$ windows in fixed impact parameter collisions. The skewness increased consistently for higher $\eta$ windows in all non-QGP models like PYTHIA, PHOJET, QGSJET, and DPMJET. The kurtosis of the distribution also increased with $\eta$ windows in non-QGP models. The skewness and kurtosis of the distributions produced by HYDJET++ decreased with $\Delta\eta$ and eventually reached zero at higher $\Delta\eta$.

\end{abstract}

\maketitle


\section{\label{sec:level1}Introduction\protect\\ }
A Quark-Gluon plasma is a hot, dense state of deconfined partons formed in the collision of heavy-ion beams at relativistic speeds \cite{Cabibbo:1975ig,Chapline:1976gy,Bjorken:1967fb}. A QGP formation is verified in an experiment by its signatures, like collective flow, jet quenching, strangeness enhancement, charmonium suppression, etc \cite{Singh:1992sp,STAR:2013ayu}. The elliptic flow that results from the transfer of spatial anisotropy into momentum anisotropy during the expansion is one of the most reliable evidence of QGP formation \cite{Voloshin:1994mz,Snellings:2011sz,Poskanzer:1998yz,Huovinen:2001cy,Huovinen:2006jp}. The elliptic flow can be calculated using the event plane method or by the use of cumulants. The simplest way to calculate elliptic flow is the two-particle cumulant method \cite{STAR:2022pfn,Borghini:2000sa,Borghini:2001vi}, i.e,

\begin{equation}
    v_2=(1/N_{pairs}\Sigma \Sigma \cos(2\Delta \phi) )^{1/2}
\end{equation}
Where $\Delta \phi$ is the azimuthal angle difference between a particle pair. The quantity $c_2\{2\}=<c_2>=1/N_{pairs}\Sigma \Sigma \cos(\Delta \phi) $ is called the two-particle cumulant, and it is the main observable in non-flow studies. We have used $c_2$ to denote the two-particle cumulant of a single event, and $<c_2>$ is used to denote the average over all events. Ideally, an event gets a non-zero $c_2$ due to hydrodynamic flow. However, a non-zero $c_2$ can be mimicked by other processes like jet correlations or decays. These non-hydrodynamic contributions do not indicate QGP formation; thus, the term non-flow is used to describe them. The non-flow contributions usually even out in high multiplicity collisions like Pb-Pb collisions at LHC or the central Au-Au collisions at RHIC. In low multiplicity collisions, the non-flow contributions are significantly high, and they can hinder the conclusion about QGP formation. Hence, understanding the nature of non-flow contributions is crucial in identifying QGP formed in low multiplicity collisions like Au-Au in BES energies or collisions like p-Pb, d-Au, p-Au, etc.

It is a common practice to average out two-particle cumulants while calculating elliptic flow. The event-by-event two-particle cumulant distribution can provide a much deeper insight into the nature of flow and non-flow contributions, as we have shown in an earlier work. There are methods like introducing $\eta$-gap and the use of higher order cumulants to remove the non-flow correlations. These methods work reasonably well in removing non-flow. The major goal of this paper is to do a detailed analysis of the statistical properties of non-flow contributions and provide an alternative to cross-check the non-flow contamination. PYTHIA and HYDJET++ are used as proxies for non-flow and flow distributions, respectively. The models like PHOJET \cite{Bopp:1998rc}, DPMJET \cite{Roesler:2000he}, EPOS-LHC\cite{Pierog:2013ria}, and QGSJET \cite{Ostapchenko:2024myl} are used occasionally to show model-dependent aspects of non-flow correlations.

This article is organized as follows: in Section II, we have provided a brief description of HYDJET++ and PYTHIA8/ Angantyr. In Section III, we have discussed the properties of two-particle cumulants in different collision systems. In Section IV, we have summarized our work.

\section{Physics framework}
\subsection{PYTHIA}
PYTHIA is an all-purpose Monte-Carlo event generator \cite{Bierlich:2016vgw,Bierlich:2022pfr,Sjostrand:2004ef,Sjostrand:2006za}. It can be used to simulate a wide range of collisions, like ee, ep, pp, and AA collisions. The PYTHIA simulation starts with the initial hard scatterings, followed by the interactions between the beam remnants, called the Multi-parton interactions \cite{Sjostrand:2017cdm}. The parton showers \cite{Mrenna:2003if} are simulated simultaneously and added back to the initial partons. Finally, the extended strings are fragmented to form hadrons, and unstable particles decay into simpler hadrons. 

Angantyr is the heavy-ion model of PYTHIA. It stacks individual NN sub-collisions together to simulate a heavy-ion event. The sub-collisions are determined using a modified Glauber model\cite{Heiselberg:1991is,Alvioli:2014eda}, which considers color fluctuations in nucleon sub-structures. After the sub-collisions are determined, they are classified into absorptive, wounded target/projectile, double diffractive, and elastic sub-collisions. The absorptive sub-collisions are treated similarly to the non-diffractive pp collisions. The wounded target/projectile sub-collisions are treated similarly to single diffractive pp collisions. The double diffractive and elastic sub-collisions are simulated accordingly. The hadronization is performed using the LUND string fragmentation model \cite{Andersson:2001yu}.

\subsection{HYDJET++}
The HYDJET++ is a Monte-Carlo event generator popularly used to simulate symmetric heavy-ion collisions\cite{Lokhtin:2009be,Lokhtin:2008xi,Lokhtin:2010zz}. The model simulates the soft thermal processes and hard processes separately. The soft part of the model is simulated using the FAST MC model \cite{Amelin:2006qe,Amelin:2007ic}, and the hard part is simulated by the PYQUEN energy loss algorithm. The initial patron spectra and jet production vertices
from individual N-N sub-collisions are generated by PYTHIA 6. A tunable parameter $p_{T}^{min}$ (minimum momentum transfer in parton-parton scatterings) controls the jet contribution from PYTHIA. The PYQUEN calculates collisional and radiational energy losses of PYTHIA-generated jets \cite{Lokhtin:2011qq}. Finally, the hadronization is performed using the LUND string fragmentation model \cite{Andersson:2001yu}.

The soft thermal state in HYDJET++ is generated on a freeze-out hypersurface with preset freeze-out conditions \cite{Bjorken:1982qr}. The HYDJET++ model considers both chemical and thermal freeze-out with ($T_{ch}>T_{th}$).  The system expands hydrodynamically with frozen chemical composition in between these two freeze-outs, then cools down, and the hadrons stream freely after thermal freeze-out $T_{th}$. The hydrodynamic flow is incorporated by modifying the freeze-out hypersurface. The azimuthal spatial anisotropy $\epsilon$ and azimuthal momentum anisotropy determine the elliptic flow in HYDJET++. The azimuthal spatial anisotropy is assumed to be related to initial spatial eccentricity, $\epsilon=k\epsilon_0$, where $\epsilon_0=b/2R_A$ and $k$ is a scaling factor. The momentum anisotropy $\delta$ is given by:

\begin{equation}
    \delta=\frac{\sqrt{1+4B(\epsilon+B)}-1}{2B},B=C(1-\epsilon^2)\epsilon
\end{equation}

The conversion of spatial anisotropy to momentum anisotropy, which is controlled by the parameter "C". In this work, we have fixed the value of $\mathrm{C=2} $, which was used at 200 GeV in the corresponding paper \cite{Lokhtin:2008xi}. The relation between the anisotropy parameters and elliptic flow is given by \cite{Wiedemann:1997cr}:
\begin{eqnarray}
v_2 \propto \frac{2(\delta - \epsilon)}{(1-\delta^2)(1-\epsilon^2)}
\label{eq:12}
\end{eqnarray}
A more detailed version of the model can be found in the corresponding papers \cite{Amelin:2006qe, Amelin:2007ic,Lokhtin:2008xi}. The model is used in two configurations: 1. default version with parameterized hydrodynamics, and 2. jet only, which only generates the events without hydrodynamics. 

HYDJET++ is not a true hydrodynamic model since it lacks the evolution stage of the fireball. Instead it uses a Bjorken-like parameterization of the freeze-out hypersurface to calculate final state particles. The use of parameters like $\epsilon$ and $\delta$ has been successful in describing the collective flow data in a wide range of collision systems and beam energies \cite{Nayak:2024jbt,Pandey:2021ofb}. Since we are focusing only on the final state particles, HYDJET++ can be used as a reliable tool.  

The PHOJET, DPMJET, and QGSJET models were run using Python-based cromo \cite{Fedynitch:2025mgj}. These models are generally used for cosmic ray physics and lack hard jets like PYTHIA. They still produce non-flow correlations due to string fragmentation and decays.

\section{Results}
\begin{figure}[h]
\includegraphics[width=.5\textwidth]{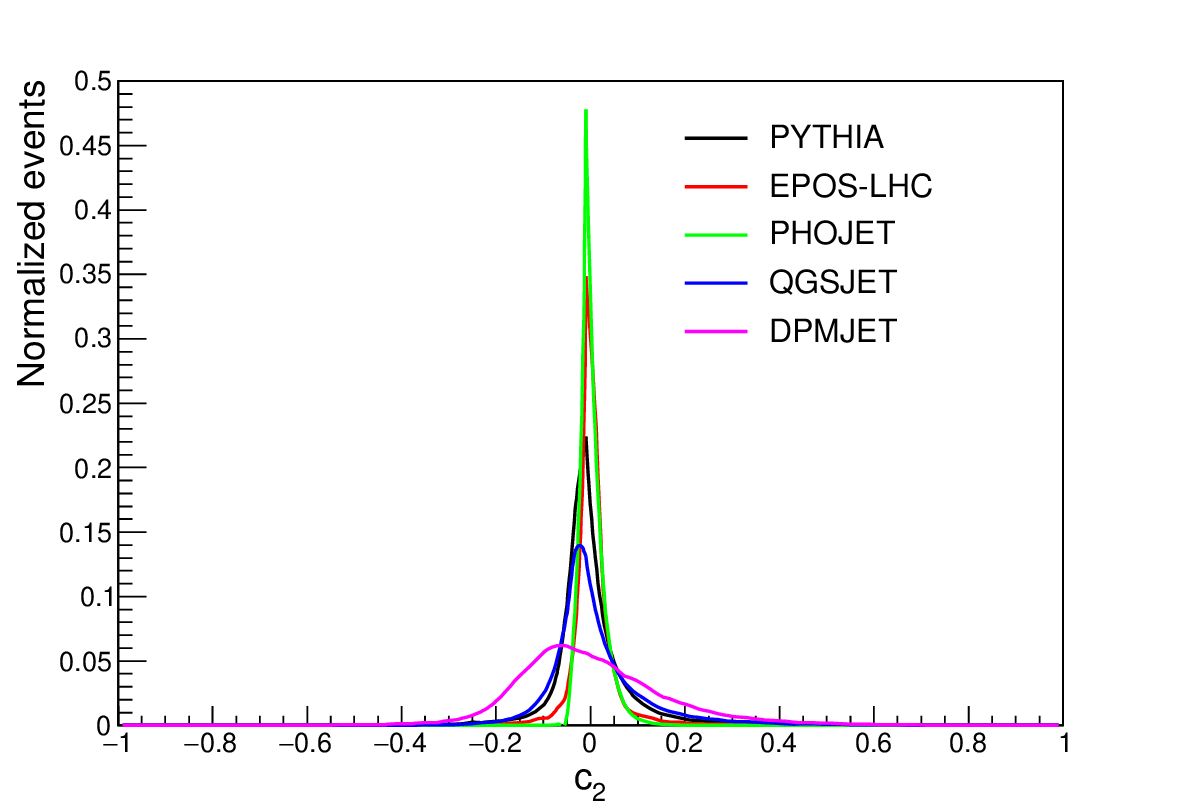}
\caption{\label{fig:1} The two-particle cumulant $c_2$ distribution of charged hadrons with different models in pp collisions at 200 GeV. }
\end{figure}
 \begin{figure}[h]
\includegraphics[width=.5\textwidth]{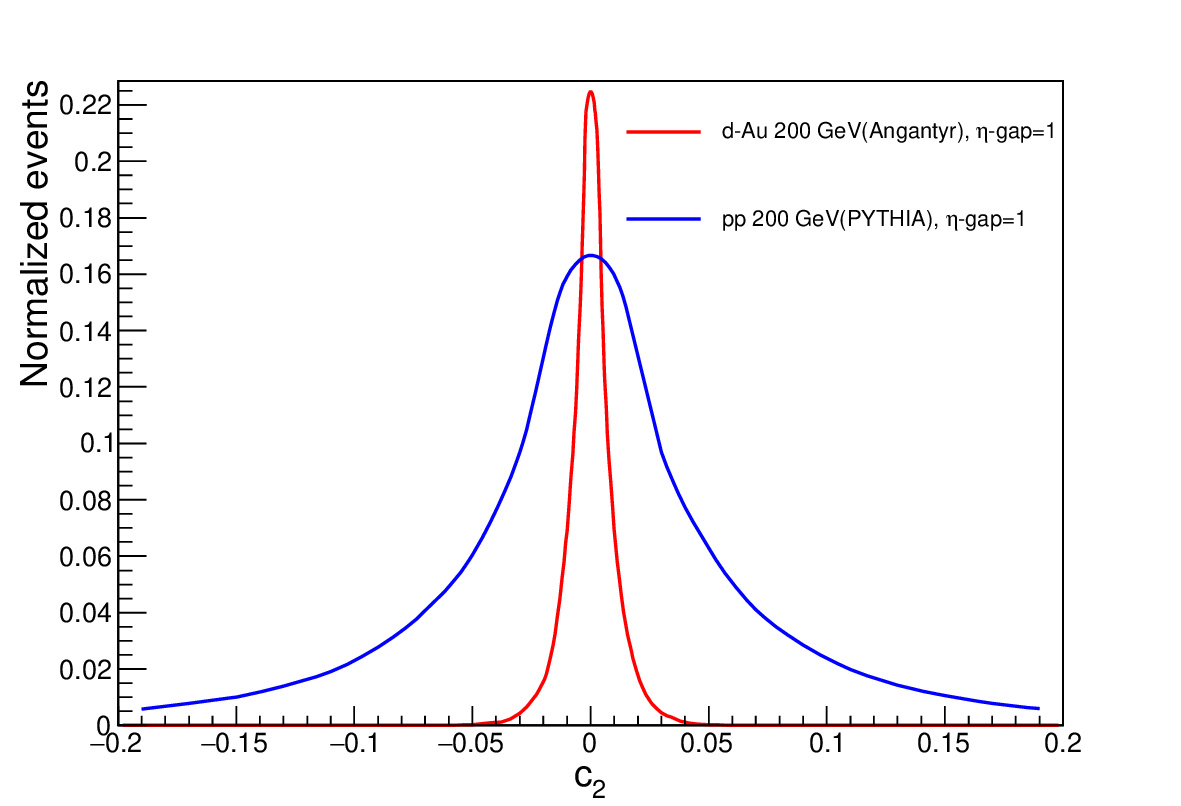}
\caption{\label{fig:1} The two-particle cumulant $c_2$ distribution of charged hadrons in d-Au collisions and pp collisions with a $\eta$-gap. }
\end{figure}

\begin{figure}[h]
\includegraphics[width=.5\textwidth]{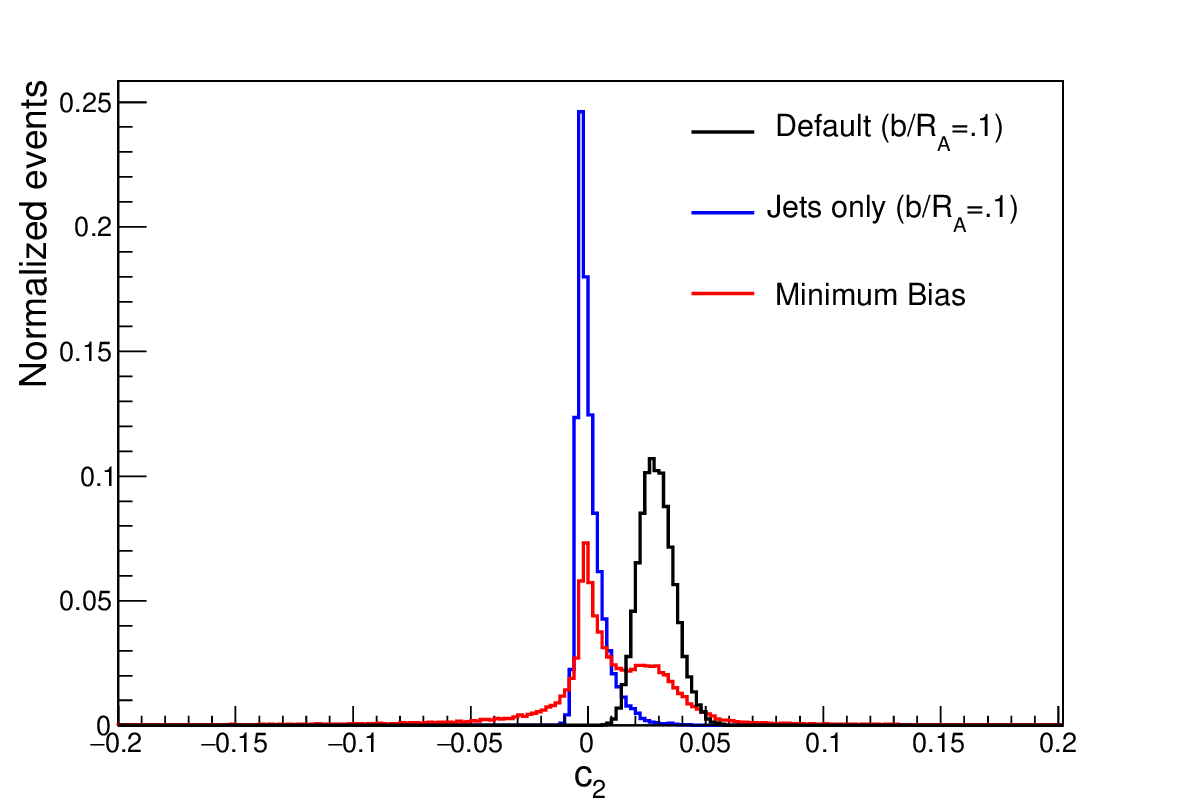}
\caption{\label{fig:2} The two-particle cumulant $c_2$ distribution of charged hadrons in Au-Au collisions at 200 GeV with different configurations of HYDJET++. }
\end{figure}
\begin{figure}[h]
\includegraphics[width=.5\textwidth]{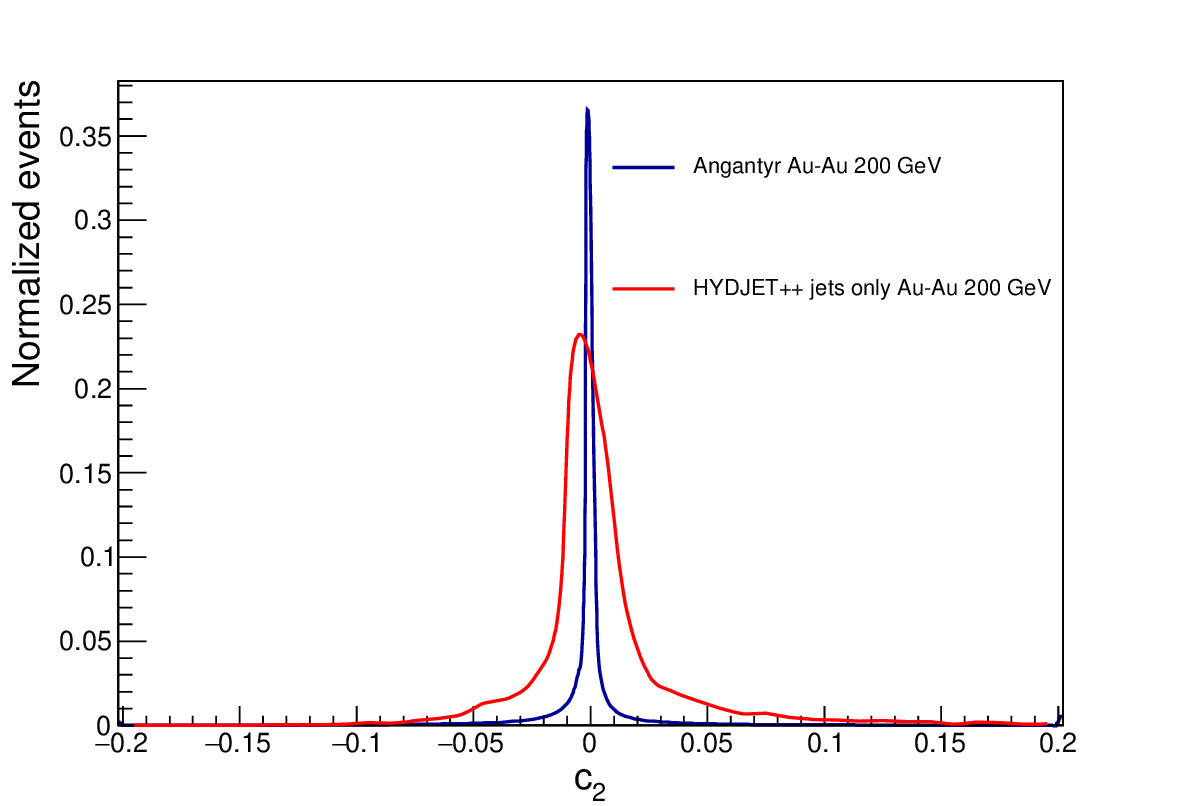}
\caption{\label{fig:3} The two-particle cumulant $c_2$ of charged hadrons in minimum bias Au-Au collisions at 200 GeV using HYDJET++ (jets only) and Angantyr. }
\end{figure}

\begin{figure}[h]
\includegraphics[width=.5\textwidth]{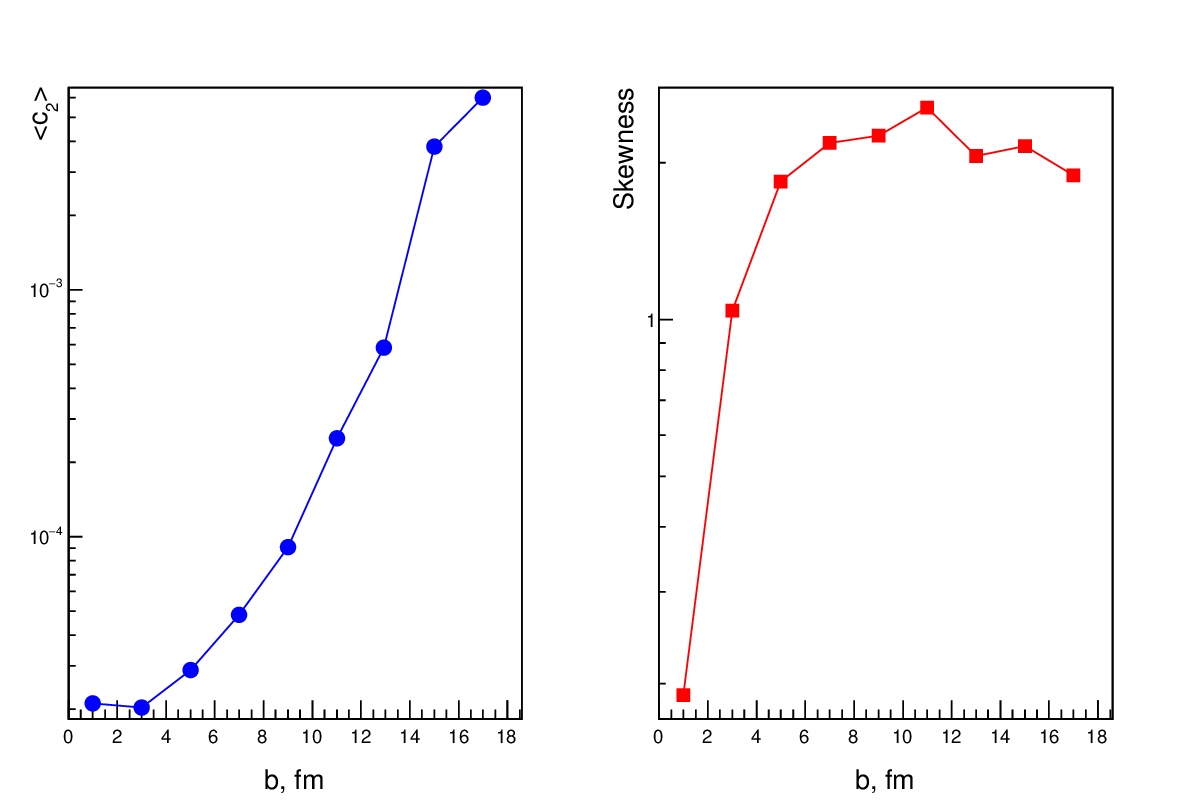}
\caption{\label{fig:3}The average two-particle cumulant $<c_2>$ of charged hadrons and the skewness of $c_2$ distribution as a function of impact parameter in Au-Au collisions using PYTHIA8/Angantyr. }
\end{figure}

\begin{figure}[h]
\includegraphics[width=.5\textwidth]{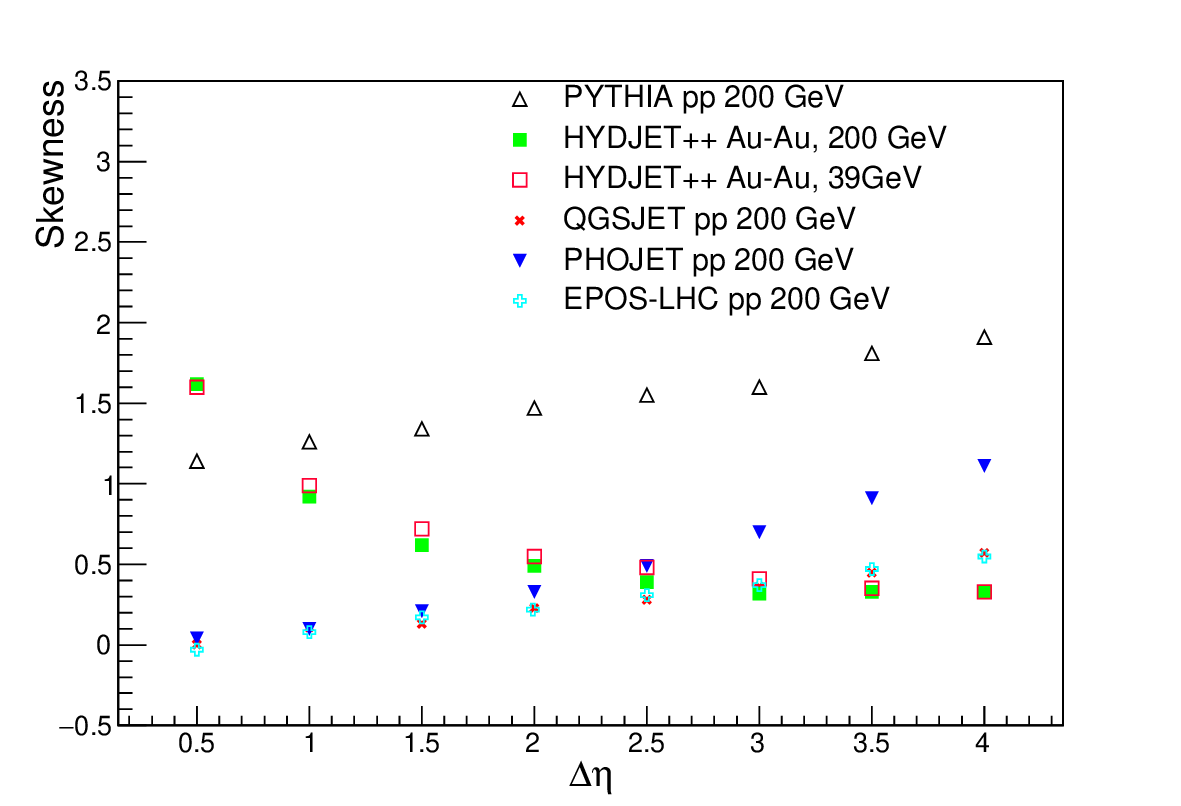}
\caption{\label{fig:3} The skewness of $c_2$ distribution of charged hadrons in pp and Au-Au collisions using different models. }
\end{figure}

\begin{figure}[h]
\includegraphics[width=.5\textwidth]{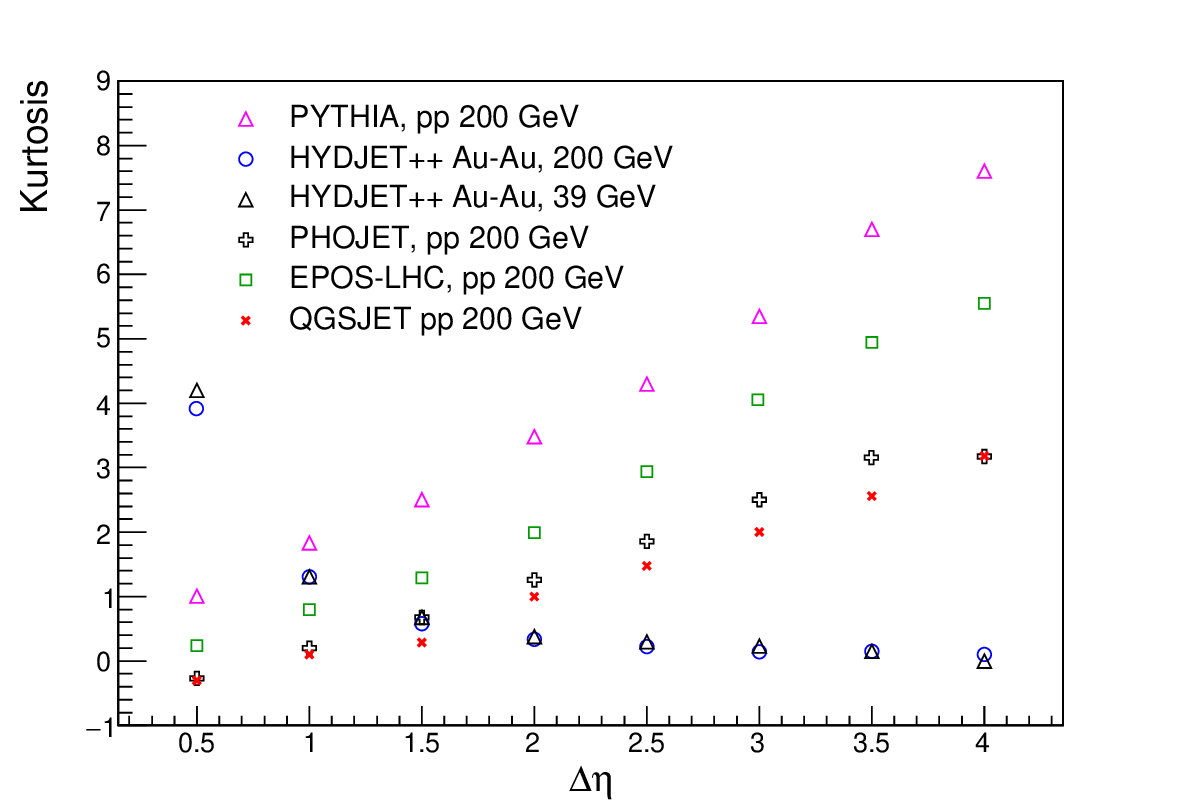}
\caption{\label{fig:3} The kurtosis of $c_2$ distribution of charged hadrons in pp and Au-Au collisions using different models. }
\end{figure}
The pp collisions at RHIC energies provide one of the best scenarios to study non-flow correlations. The pp collisions have a significantly smaller system size compared to a heavy ion collision, and they are highly unlikely to form a QGP. Unfortunately, due to the unavailability of two-particle cumulant distribution data from experiments, we will limit our discussion to simulated datasets. In Fig. 1, we have shown the two-particle cumulant distribution of charged hadrons in pp collisions at 200 GeV in the -4$<\eta<$4 pseudorapidity window using PYTHIA, EPOS-LHC, PHOJET, QGSJET, and DPMJET. It can be seen that the key attributes of distributions like mean and peak are model-dependent, but all models formed a positively skewed distribution. This aspect of the two-particle cumulant distribution is model independent. As per the central limit theorem(CLT) \cite{CentralLimitTheorem:2008,Pollard:1982}, if the sample is: 1) significantly large (number of individual pairs), 2) each contribution ($ \cos 2\Delta\phi$) is independent, 3) the mean and variance are finite, then the probability distribution of the mean ($<\cos2 \Delta\phi>$) of such samples approximates a normal distribution. For instance, let's consider the two-particle cumulant distribution of charged hadrons calculated using a $\eta$-gap. The introduction of $\eta$-gap significantly suppresses the non-flow correlations, like jets and decays. The individual $\cos\Delta\phi$s are now independent and random; their probability distribution approximates a Gaussian. Although the distribution is not a perfect Gaussian due to non-zero excess kurtosis, the distribution remains symmetric and lacks the skewed tail. A clear illustration of $c_2$ distribution of charged hadrons in d-Au collision with a $\eta$-gap=1 ($N_{ch}>120$ in -4 to 0 pseudorapidity range) can be found in Fig. 2. The two sub-events were selected in -4 to 0 and 1 to 5 pseudorapidity windows, respectively. A similar pattern can also be observed in pp collisions.  The two-particle cumulant distribution is not completely random. The events with a few jet correlations or decays tend to have higher positive values, thus biasing the distribution and forming the signature skewed tail. The tail is most prominent in PYTHIA due to the contribution from processes like jets, mini-jets, MPIs, and string fragmentations. The models like DPMJET, PHOJET, and QGSJET, which are popularly used for cosmic ray studies, do not have hard jets like PYTHIA. So, the tails of the distributions obtained using such models are not as pronounced as those of PYTHIA. The self-correlations can also produce similar skewed distributions. So, special attention was given while computing $c_2$ values. In the current study, we have not considered pairs with the same $\phi$s ($\phi_i=\phi_j$). Only the pairs with $\phi_i\neq\phi_j$ were used while calculating $c_2$ values. 

In HYDJET++,  the two-particle cumulant correlates with initial spatial anisotropy. The initial eccentricity of a collision fluctuates on an event-by-event basis due to fluctuations in the number and transverse position of quarks in the colliding nuclei \cite{Voloshin:2006gz}. It was pointed out by Voloshin et. al. in 2007 that the event-by-event spatial anisotropy distribution assumes a Gaussian form \cite{Voloshin:2006gz,Voloshin:2007pc}. These fluctuations are transferred into the final state azimuthal anisotropy and the two-particle cumulants. In HYDJET++, a Gaussian fluctuation term is added to the final state spatial anisotropy to incorporate these fluctuations. So, the two-particle cumulant distribution generated using HYDJET++ should be a normal distribution if the impact parameter is fixed. Each impact parameter corresponds to a unique initial spatial anisotropy, i.e $\epsilon_0=b/2R_A$, around which the $\epsilon_0$ values form a Gaussian due to event-by-event fluctuations. The Gaussian shape of the $c_2$ distribution in HYDJET++ originates due to the event-by-event fluctuation of spatial anisotropy, not due to the Central Limit Theorem, as the $\cos2\Delta\phi$s are not independent and random. If a range of impact parameters is chosen, there are several values of $\epsilon_0$ corresponding to each value of impact parameter. So, the distribution assumes a plateau-like shape due to the superposition of several Gaussians. In Fig. 3, we have shown the two-particle cumulant distribution of charged hadrons in Au-Au collisions at 39 GeV for impact parameter $b=0.1/R_{A} $. The default configuration forms a Gaussian, while the configuration with "Hydro off" forms a positively skewed distribution. As discussed earlier, the minimum biased collisions do not form any distinct distribution due to the lack of any constant value of spatial anisotropy ($\epsilon_0$). The non-flow correlations are independent of impact parameter and spatial anisotropy. So, they produce similar distributions both in fixed impact parameter and minimum-biased collisions. The $c_2$ values of charged hadrons in minimum bias Au-Au collisions at 200 GeV using HYDJET++ (jets only) and Angantyr can be found in Fig. 4. It should be noted that the event-by-event $c_2$ distribution depends on how a model treats the initial state fluctuations and can be model dependent.

The two-particle cumulant distribution does not produce any standard distribution. So, we have used their deviation from the Gaussian for this analysis. The deviation from a normal distribution (Gaussian) is characterized using skewness and excess kurtosis. The skewness describes the skewed deviation of the distribution from the mean value. The kurtosis describes the peakedness or the tailedness of the distribution. 

\begin{equation}
    sk=\frac{\Sigma_{i}^{N} (x_i -<x>)}{(N-1)\sigma^3}
\end{equation}
\begin{equation}
    kr=\frac{1/N\Sigma_{i}^{N}(x_i-<x>)^4}{(1/N\Sigma_{i}^{N}(x_i-<x>)^2)^2}-3
\end{equation}

 Where $x_i$ is the random variable, $<x>$ is the mean, and $\sigma$ is the standard deviation.

In Fig. 5, we have shown the $<c_2>$ of charged hadrons in Au-Au collisions using PYTHIA8/Angantyr for different impact parameters. The $<c_2>$ rises rapidly with impact parameter due to decreasing multiplicity. The skewness of the $c_2$ distribution also increases with the impact parameter and remains almost constant in peripheral collisions. It can be claimed that a higher value of $<c_2>$ corresponds to higher skewness. In Fig. 6, we have shown the skewness of the two-particle cumulant distribution of charged hadrons as a function of pseudorapidity window in pp collisions using PYTHIA, PHOJET, QGSJET, and EPOS-LHC. The distributions show non-zero skewness regardless of the model used. The skewness increases with $\Delta\eta$ windows in all models. This can be attributed to the increased number of non-flow correlations at higher $\eta$ windows. The skewness of the $c_2$ distribution from HYDJET in Au-Au collisions at 200 GeV and 39 GeV (b=0.1 $R_A$) is shown in the same figure. The skewness of the $c_2$ distribution in d-Au collisions also follows a similar trend\cite{Nayak:2025exz}. Unlike the non-hydrodynamic models, the skewness decreases with $\Delta\eta$, eventually reaching zero at higher $\Delta\eta$. This suggests that the HYDJET++ model produces a smooth Gaussian distribution at higher $\Delta\eta$ windows. 

In Fig. 7, we have shown the kurtosis of the $c_2$ distribution in pp collisions using PYTHIA, PHOJET, QGSJET, and EPOS-LHC. Similar to skewness, the kurtosis remains higher than zero for all models. However, the kurtosis does not originate from the non-flow correlations, as the $\eta$-gaps introduced earlier did not suppress kurtosis. The individual $c_2$ values are the average of $\cos\Delta\phi$ values. If the number of random pairs increases, the average value approaches zero. This makes the overall distribution stiffer and increases the kurtosis. This explains the stiffer distribution produced in d-Au collisions compared to pp collisions in Fig. 2. The kurtosis increases with $\Delta\eta$ in all non-QGP models due to the increase in the number of random $\Delta\phi$ values. The kurtosis of the $c_2$ distribution generated from HYDJET++ behaves opposite to the ones generated using non-QGP models. The kurtosis decreases with $\Delta\eta$ and approaches zero at higher $\Delta\eta$, similar to skewness. The $\cos\Delta\phi$ values are a result of hydrodynamic flow, and they are not randomly correlated. So, the kurtosis from HYDJET++ does not increase with $\Delta\eta$. The $c_2$ distribution of charged hadrons as a function of $\Delta\eta$ in Au-Au collisions at 200 GeV and 39 GeV (b=0.1$R_A$) can be found in Fig. 7.

In summary, the skewness of the $c_2$ distribution is a result of non-flow correlations like jets and decays, and the kurtosis originates from the remaining random correlations. 

\section{Summary}
 The paper reports the event-by-event behavior of the two-particle cumulant distribution of charged hadrons in pp and heavy-ion collisions using different theoretical models. The quantities like the mean and variance of the $c_2$ distribution are model-dependent. However, the distributions generated using non-QGP models like PYTHIA, PHOJET, QGSJET, and DPMJET produced a signature positively skewed tail in pp collisions at 200 GeV. The large $\cos 2\Delta\phi$ values originating from the non-flow correlations like jets and decays are the primary cause of the skewed tail. The introduction of $\eta$-gap suppresses the non-flow correlations and removes the skewed tail.

The event-by-event eccentricity fluctuations have a Gaussian smearing. These fluctuations transfer to the final state anisotropies as well as to the two-particle cumulants. The HYDJET++ produced a Gaussian distribution in Au-Au collisions due to these eccentricity fluctuations. The Gaussian distribution was only observed in fixed-impact parameter collisions with hydrodynamics. The minimum-biased collisions and simulations without hydrodynamics did not produce a normal distribution. The skewness of the $c_2$ distributions obtained using non-QGP models increased steadily with $\Delta\eta$ windows due to the increase in the number of non-flow correlation pairs. The kurtosis also increased with $\Delta\eta$ due to the enhancement in the number of random $\cos 2\Delta\phi$ values that made the distribution narrower and stiffer, increasing the kurtosis. On the contrary, the skewness and kurtosis of the distributions obtained using HYDJET++ decreased with $\Delta\eta$ and approached zero at higher $\Delta\eta$. This suggests that the two-particle cumulant distribution deviates from Gaussian and forms a signature skewed tail in the presence of non-flow correlations. The distribution approaches a Gaussian with negligible skewness and kurtosis when hydrodynamics is included. This aspect of the $c_2$ distribution can be used as an alternative to cross-check non-flow contamination in smaller collision systems.
 
\begin{acknowledgments}
BKS sincerely acknowledges financial support from the Institute of Eminence (IoE) BHU Grant number 6031. SRN acknowledges the financial support from the UGC Non-NET fellowship and IoE research incentive during the research work.  AD acknowledges an institute fellowship from PDPM Indian Institute of Information Technology Design \& Manufacturing, Jabalpur, India.

\end{acknowledgments}


\end{document}